\documentstyle[epsfig]{elsart}

\begin{document}

\begin{frontmatter}

\title{Anatomy of three-body decay \\
I. Schematic models}

\author{E. Garrido} 
\address{ Instituto de Estructura de la Materia, CSIC, 
Serrano 123, E-28006 Madrid, Spain }

\author{D.V. Fedorov, A.S.~Jensen \and H.O.U. Fynbo}
\address{ Department of Physics and Astronomy,
        University of Aarhus, DK-8000 Aarhus C, Denmark }

\date{\today}

\maketitle

\begin{abstract}
Sequential three-body decay proceeds via spatially confined
quasi-stationary two-body configurations.  Direct three-body decay
populates the three-body continuum without intermediate steps.  The
relative importance of these decay modes is discussed in a schematic
model employing only Coulomb or centrifugal barrier potentials.
Decisive dimensionless charge, mass and energy ratios are derived.
Sequential decay is usually favored for charged particles.  Small
charge and small mass of high energy is preferably emitted first.
Without Coulomb potential the sequential decay is favored except when
both resonance energy and intermediate two-body energy are large.
\end{abstract}

\end{frontmatter}

\par\leavevmode\hbox {\it PACS:\ } 21.45.+v, 31.15.Ja, 25.70.Ef

\maketitle

\section{Introduction.}

Resonances and excited states decaying into two clusters have been
thoroughly studied from the early days of quantum mechanics.  The
prominent examples are $\alpha$-decay, nucleon emission and fission,
see e.g. \cite{sie87}.  More complicated final states with many
fragments also occur and in particular three-body decay has been
studied on and off over many years, see e.g. \cite{goo65}.  Many
recent investigations focused on the unusual nuclear three-body halo
structures which are naturally inclined to decay into their
constituent clusters \cite{oza01,jen04}.  General scaling properties
for direct multi-cluster Coulomb decay were also recently derived
\cite{kar04}.

The primary experimental decay information is the distribution of
cluster energies after the decay. From these spectra the decay can be
analysed as sequential decay, i.e. emission of one particle populating
a subsequently decaying resonance of the two remaining particles, or
direct decay into the three-body continuum.  Early attempts were made
to characterize nuclear three-body decay \cite{dan87,boc93}.  Here the
sequential decay is two subsequent two-body decays. The direct decay
is parametrized by one or two terms of an expansion in hyperspherical
harmonics. This may be sufficient for short-range interactions
although even then the crucial asymptotic behavior should be
influenced by admixtures of higher harmonics arising via the structure
of the decaying resonance. In contrast a few hyperharmonics are most
likely inadequate for charged particles in the final state.

The experimental techniques improved tremendously over the last
decade.  Kinematically complete and accurate experimental information
becomes available.  The three-body decay experiments are essentially
all consistent with sequential decay with emission of one particle at
a time.  The observables available are widths of the decaying
resonances and angular distributions and energy spectra of the emitted
particles, see e.g. references in \cite{gar05}.

In quantum mechanics all paths connecting initial and final states
contribute to the decay width.  However, the least action path often
dominates and provides sufficient accuracy.  For given masses the
least action path can be determined from suitable potential energy
surfaces covering both initial and final states. The width can then be
calculated and the decay mechanism related to tunneling properties of
the dominating potentials.  In this paper we assume that an initial
state is a many-body resonance decaying into three fragments.  We
model this as a three-body system at intermediate and large distances.
The three particles in the final state are then formed before entering
the barrier precisely as the preformation factor in the description of
$\alpha$-decay. This invites to definitions (and subsequent
experimental determination) of three-body spectroscopic factors
describing the fraction of corresponding three-body content in the
initial many-body wave function as previously attempted for
$\alpha$-cluster states \cite{ich73}.

The purpose of the present paper is to investigate the 
mechanism for three-body decay employing only Coulomb and centrifugal
barriers.  This is a generalization of classical $\alpha$-decay
calculations to analogous three-body tunneling computations.  We shall
in particular concentrate on differences between sequential and direct
decay.  Realistic computations including both short and long-range
potentials are discussed in the companion paper \cite{gar05}.

\section{Theoretical concepts}

The hyperspherical adiabatic expansion method is often quantitatively
fairly accurate with only the dominating potential \cite{nie01}.  The
corresponding generalized effective radial potentials are expressed as
function of the hyperradius, $\rho$, defined by
\begin{equation} \label{e20}
  \rho^2 \equiv  \frac{1}{m M}  \sum_{i<k} m_i m_k 
 {\bf r_{ik}}^2  =  ({\bf x_j}^2 + {\bf y_j}^2) \;\; , \;\;
 {\bf r_{ik}}^2 =  ({\bf r_i} - {\bf r_k})^2 \; ,
\end{equation}
where $(i,j,k)$ is a permutation of $(1,2,3)$, ${\bf r_i}$ is the
coordinate of particle $i$, $M = \sum m_i$ and $m$ is an arbitrary
normalization mass, ${\bf x_j}$ and ${\bf y_j}$ are respectively
proportional to the distance between two particles and the distance
between their center of mass and the third particle. The remaining
coordinates are all angles \cite{nie01}, but they are not necessary in
the present paper.

For short-range interactions the lowest hyperspherical adiabatic
potentials dominate in the expansion of the wave function.  The
efficiency is reflected by the ability to describe the Efimov effect
arising when two subsystems simultaneous have large scattering lengths
\cite{fed94}.  For systems with both short and long-range interactions
the lowest potentials also dominate at small (and intermediate)
distances whereas more components usually are needed for
asymptotically large distances.  

We shall assume that one adiabatic potential is sufficient in analogy
to the two-body problem where short-range, Coulomb and centrifugal
barrier terms provide an effective one-dimensional radial potential
with possible bound states and resonances.  With the dominating
adiabatic potential the width of a given three-body resonance can be
estimated by the WKB tunneling probability multiplied by the knocking
rate \cite{sie87}.  This is a conceptual extension from two to
three-body decay which is far from being obvious but shown recently to
be approximately valid \cite{fed02}.

In principle different paths lead from the initial resonance state
located at small distances to the final free three-particle state.
Large separation between all pairs of particles can for example be
achieved in two steps, i.e. first by moving one particle to infinity
while the other two remain at essentially the same distance from each
other and second by moving the two close-lying particles apart.  Let
us define sequential decay as this two-step process if the second step
is started after the first particle is at a distance larger than the
initial size of the three-body system.

The intermediate configuration does not have to be a two-body
resonance. A substantial attraction for example in $s$-waves arising
from a close-lying virtual $s$-state could be sufficient to produce
the signature of such a sequential decay. This is in close analogy to
the effects of final state interactions in fragmentation reactions of
three-body systems where two final-state fragments remain close
together resulting in a significantly narrower momentum distribution
\cite{gar96}.

To clarify the idea we assume that the third particle is emitted with
relative energy $E_{12,3}=E-E_{12}$, where $E$ is the total energy and
$E_{12}$ is the energy of the remaining two-body resonance.  The
corresponding velocity is $v_{12,3}=(2E_{12,3}/\mu_{12,3})^{1/2}$,
where $\mu_{12,3}$ is the related reduced mass. The two-body resonance
has a lifetime of $t_{12}= \hbar/\Gamma_{12}$. The distance, $d_{12,3}$,
particle 3 moves before the $12$-system decays compared to the size of
the three-body system $R_0$ is then
\begin{equation} \label{e40}
 \frac{d_{12,3}}{R_0} = \frac{v_{12,3} t_{12}}{R_0} = 
 \frac{1}{\Gamma_{12}} \sqrt{\frac{2\hbar^2(E-E_{12})}{\mu_{12,3}R_0^2}}
 \;  ,
\end{equation}
which should be larger than one to fulfill the conditions for
sequential decay. Typical numbers for nuclei gives that the two-body
decay width $\Gamma_{12}$ then has to be smaller than about 1~MeV.
However, this condition is not sufficient to decide the character of
the decay.  The picture is not useful when the initial three-body
resonance wave function has no configuration similar to a two-body
subsystem in a resonance.  Then direct decay has an advantage.  The
over all conclusion is that the preference for sequential decay is
indicated by a narrow two-body resonance, but this is only a necessary
and not a sufficient condition.

\section{Schematic models}

Let us assume the fragments are formed at distances smaller than or
equal to the inner classical turning point of the barrier defined by
the dominating adiabatic potential. This is analogous to
$\alpha$-decay with a preformation factor equal to one.  If this
clusterization of the final state fragments is not fully present a
pre-exponential preformation factor has to be included.  The
terminology is identical to the description of $\alpha$-decay and our
formulation is a direct generalization to three particles in the
final state.

To discuss direct versus sequential decay we want to compare the
corresponding decay probabilities.  Assume that the sequential process
occurs by first emitting one of the particles and subsequently the
remaining two-body system decays.  The boundary condition of an
outgoing flux means that we should compare the direct decay width with
the width for the first step sequential decay multiplied by the
branching ratio for decaying into the final state by the next step.
The second step may be in competition with other decay modes. This is
the classical picture where a decision of paths is chosen both
initially and in the next step.  Thus the sequential decay width is
the width for the first step multiplied by the branching ratio for the
second step, i.e. the second width divided by itself added to the
total width for decay into competing channels.

It is rewarding to reconcile this procedure with the sequential decay
condition of small partial two-body width $\Gamma_{12}$ arising from
Eq.(\ref{e40}).  If $\Gamma_{12}$ is large the total sequential decay
width is given by $\Gamma_{12,3}$ arising from the first step.
However, Eq.(\ref{e40}) then also indicates direct decay.  For large
$\Gamma_{12}$ other modes are strongly coupled to the sequential decay
channel.  The system does therefore not survive long enough for the
third particle to move outside the radius of the initial three-body
system.  The other channels, directly populating continuum states,
take all the probability before even the first step in the sequential
decay process is completed.  This means that the direct decay seems to
be very likely but also sequential decay may be comparably large
depending on the size of $\Gamma_{12,3}$.  Thus large $\Gamma_{12}$
may favor direct decay perhaps coupling to other adiabatic
channels. This cannot be described by one-channel estimates.

\subsection{The WKB approximation for an effective hyperradial potential}

As soon as one effective potential is responsible for the decay we can
derive simple estimates by use of the WKB tunneling transmission $T$,
i.e.
\begin{equation} \label{e50}
 T = \frac{1}{1+ \exp(2S)} \approx \exp(-2S) \; \; , \;\; 
 S = \frac{1}{\hbar} \int_{\rho_0}^{\rho_t} {\rm d}\rho
\sqrt{2 m (V (\rho)-E)} \;  ,
\end{equation}
where $E$ is both the total energy and the kinetic energy of the
particles after separation, $\rho_0$ and $\rho_t$ are the classical
turning points where $V(\rho_0) = V(\rho_t) = E$. The integration path
here is along $\rho$ but the expression remains valid for any other path.
This perhaps unusual expression for $T$ is really the second order WKB
approximation which substantially extends the validity range
\cite{sie87}. For example the harmonic oscillator transmission coefficient 
is then exactly reproduced.  The value of $T=1/2$ obtained at the top
of the barrier ($\rho_0 = \rho_t, S= 0$) is quantum mechanically
correct.  When $S\gg 1$ the usual exponential expression in
Eq.(\ref{e50}) is obtained.

A rather good estimate is found with the dominating adiabatic
effective hyperradial potential $U_{eff}$ inserted instead of $V$, see
e.g. \cite{fed02}.  A number of crucial effects are collected in
$U_{eff}$, e.g. the three distinguishable contributions from
centrifugal barrier, Coulomb and short-range potentials, and the
subtle adiabatic adjustments of structure as the distance between
clusters increases.  Then intermediate structures are picked up along
the decay path, e.g. a particular two-body resonance or an especially
strong $s$-wave attraction between two particles.  Furthermore,
several configurations can be present and interfere to produce
$U_{eff}$, e.g. different resonances in the two-body subsystems or the
coherent contributions arising due to (anti)symmetry of identical
particles.  All these effects are accounted for since each adiabatic
potential represents a very specific weighted combination of paths
continuously leading from small to large distances.

Different adiabatic potentials represent different paths which
separately can be estimated by the WKB approximation. Effects of
several simultaneously contributing adiabatic potentials would one way
or another require coupled channel calculations.  Thus it is usually
not sufficient to assume a given classical path and compute the
expectation values of the different terms in the three-body
Hamiltonian.  However, as for $\alpha$-decay, it is very illuminating
to exhibit the effects of the most prominent terms in the classical
picture.  We shall therefore explicitly investigate Coulomb and
centrifugal barrier terms.  We then only implicitly include the
short-range interaction by choices of geometric paths, of two-body
resonance energies and relative angular momenta.  The same philosophy
was recently applied to direct decay for the Coulomb potential
\cite{kar04}.

\subsection{Geometries}

A given decay path is defined by specifying how the distance between
particles increase as function of $\rho$.  We assume a constant
scaling although any function in principle could be used. With the use
of Eq.(\ref{e20}) we then define the positive scaling constants
$s_{ik}$ by
\begin{equation} \label{e70}
\frac{{\bf r_{ik}}^2}{ \rho^2} \equiv s^2_{ik}  \;\; ,\;\;
 m M  \equiv    \sum_{i<k} m_i m_k s^2_{ik}  \;\; ,\;\;
   s^2_{ik} = \frac{mM}{\sum_{i<k} m_i m_k }  \;  ,
\end{equation}
where the upper limit are given by $s_{ik} < \sqrt{mM/(m_im_k)}$ as
seen from Eq.(\ref{e70}) since where each term must be smaller than
the sum $mM$. The last expression for $s_{ik}$ is obtained by assuming
that all $s_{ik}$ are identical.

\begin{figure}
\begin{center}
\vspace*{-1.1cm}
\epsfig{file=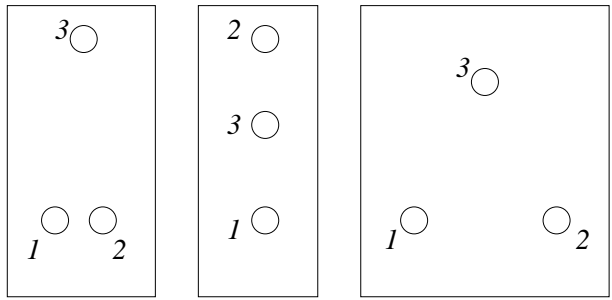,scale=0.4, angle=0}
\end{center}
\caption{ Different geometric configurations illustrating possible decay
paths, i.e. removal of one particle while the other two remain close
(left) perhaps in a two-body resonance, scaling of a linear
configuration (middle), and an overall proportional scaling of all
distances (right).  Cyclic permutations are also allowed. }
\label{fig2}
\end{figure}

A number of different paths are now parametrized by choices of
$s_{ik}$ as illustrated in Fig.\ref{fig2}.  When for example $s_{12}
\approx 0$ and  $s_{13} \approx s_{23}$ we obtain the typical sequential 
path where particles $1$ and $2$ stay close until the distance to
particle $3$ is much larger than the initial size of the three-body
system.  After this emission of particle $3$, also particles $1$ and
$2$ increase their mutual distance until all particle distances are
large, i.e. the final three-body decay has been completed.

When $s_{13} \approx s_{23} \approx s_{12}/2$ the decay proceeds
through the linear chain also depicted in Fig.\ref{fig2}.  Direct
decay to the three-body continuum is described by a simultaneous
increase of all $s_{ik}$ until all particle distances are large
compared to the initial size.  Then no subsystem is used as an
intermediate stepping stone and no preference for two-body
substructures are exploited.  Cyclic permutations or renaming of the
particles then cover all extreme structures.

\subsection{Coulomb potential}

When long-range repulsive Coulomb interactions are present they
strongly influence the decay.  We assume that the effect of the
short-range interaction is vanishingly small in the classically
forbidden barrier region. Unless the charges are very small or the
angular momenta are large the (generalized) centrifugal barrier also
has relatively small effect on the barrier penetrability \cite{sie87}.
Thus we use Eq.(\ref{e50}), either for direct decay with $V$ defined
as
\begin{eqnarray} \label{e80}
 V(\rho) &=& \sum_{i<k}  \frac{Z_i Z_k e^2}{r_{ik}} = 
 \frac{1}{\rho} \sum_{i<k}  \frac{Z_i Z_k e^2}{s_{ik}}  \;  ,
\end{eqnarray}
or with $\rho$ and $m$ substituted by $r_{13}$ or $r_{12}$ and the
related masses $\mu_{12,3}$ and $\mu_{12}$, i.e.
\begin{eqnarray} \label{e90}
 V(r_{13}) &=& E_{12} + \frac{Z_1 Z_3 e^2}{r_{13}} + \frac{Z_2 Z_3 e^2}{r_{23}}
 =  E_{12} +  \frac{(Z_1+ Z_2) Z_3 e^2}{r_{13}}  \; , 
 \\ \label{e94}
 V(r_{12}) &=& \frac{Z_1 Z_2 e^2}{r_{12}}  \; , 
\end{eqnarray}
corresponding to the two steps of the sequential decay process when
particle $3$ is emitted. The constant energy $E_{12}$ is tied up in
the intermediate configuration with particles $1$ and $2$ close
together.  This two-body structure could correspond to a resonance or
maybe result from an $s$-wave attraction keeping the particles
together.
 
The WKB exponents are in any case proportional to a generic function
like
\begin{eqnarray} \label{e95}
 S \propto   \int_{x_0}^{x_t} {\rm d} x \sqrt{\frac{x_t}{x} - 1}
 =  x_t \Big(\arctan{\sqrt{\frac{x_t}{x_0} - 1}} - 
 \frac{x_0}{x_t} \sqrt{\frac{x_t}{x_0} - 1}  \Big)
 \approx   \frac{\pi  x_t}{2}  \;  ,
\end{eqnarray}
where we assumed that $x_0 \ll x_t$.  This approximation implies that
the exponent is fairly accurately determined, but the decay width
itself is substantially more inaccurate.

Using Eqs.(\ref{e50}), (\ref{e80}), (\ref{e90}) and (\ref{e95}) we
then arrive at the WKB exponents for the direct and the two-step
sequential decays, i.e.
\begin{eqnarray} \label{e100}
  S &=& \frac{\pi}{2} \sum_{i<k}  \frac{ Z_i Z_k e^2}{ s_{ik}}  
 \sqrt{\frac{2 m  }{\hbar^2 E}}  \; = \;   
\frac{\pi}{2} \Big(\sum_{i<k}  Z_i Z_k e^2\Big)  
 \sqrt{\frac{2 \sum_{i<k}  m_i m_k     }{\hbar^2 E M}} \; ,\\ \label{e110} 
  S_{12,3} &=& \frac{\pi}{2}  (Z_1 + Z_2) Z_3 e^2 
\sqrt{\frac{2 \mu_{12,3} }{\hbar^2 (E-E_{12})}} \equiv
 \frac{b_{12,3}}{ \sqrt{E-E_{12}}}  \;  ,  \\ \label{e120}
  S_{12} &=&  \frac{\pi}{2}  Z_1 Z_2 e^2 
 \sqrt{\frac{2 \mu_{12}  }{\hbar^2 E_{12}}} \equiv
   \frac{b_{12}}{ \sqrt{E_{12}}} \;  ,
\end{eqnarray}
where the notation is self-explanatory. We used the appropriate reduced
masses ($\mu_{12},\mu_{12,3}$) for the sequential decay corresponding
to particles $1$ and $2$, and their center of mass and the third
particle, respectively.  

The last expression in Eq.(\ref{e100}) is obtained by using identical
scaling, i.e. all $s_{ik}$ are the same as expressed in
Eq.(\ref{e70}). Another set of scaling parameters can be obtained for
the potential in Eq.(\ref{e80}) by minimizing the action integral $S$
in Eq.(\ref{e50}) with respect to $s_{ik}$.  With the constraint on
$s_{ik}$ in the middle equation of Eq.(\ref{e70}) we obtain the least
action result
\begin{equation} \label{e124}
  S_{min} = \frac{\pi e^2}{2}  \sqrt{\frac{2  }{\hbar^2 E M}} 
 \Big( \sum_{i<k} (Z_i Z_k)^{2/3}   (m_i m_k)^{1/3} \Big)^{3/2}
   \equiv \frac{b}{\sqrt{E}}  \; .   
\end{equation} 
The corresponding optimum path is defined for scaling parameters given
by $s_{ik}^{3} m_i Z_j = s_{jk}^{3} m_j Z_i$.  Thus the action is
minimized with the ratio of distances between particles $1-3$ and
$2-3$ given as the cubic root of the ratio of charges $Z_1/Z_2$
multiplied by the ratio $m_2/m_1$.  For identical particles the path
is scaling of an equal sided triangle and given by the upper limit in
Eq.(\ref{e70}).  The same scaling property was also found in
\cite{kar04} by use of time dependent equations of motion for the
scaling parameters.

The direct to sequential branching ratio is a sum of two terms,
\begin{eqnarray} \label{e129}
 P_{d,seq} = \frac{T}{T_{12,3}}  \Big(1 + \frac{T_{\gamma}}{T_{12}}\Big) \; ,
\end{eqnarray}
each related to the first and second steps of the process, respectively.
From Eq.(\ref{e50}) we have that $T = \exp(-2S_{min})$, $T_{12,3} =
\exp(-2S_{12,3})$, $T_{12} = \exp(-2S_{12})$, and $T_{\gamma}$ is 
proportional to the decay probability into other modes in the second
step of the sequential decay process. In most cases of interest we
have $T_{\gamma} \ll T_{12}$ and the second term in Eq.(\ref{e129})
can be ignored.  Thus, when $T<T_{12,3}$ the sequential decay is
preferred and vice versa.  This inequality is controlled by the
relative size of $S_{12,3}$ and $S_{min}$. If we also measure $S_{12}$
relative to $S_{min}$ we have two crucial quantities appearing in
these expressions, i.e.
\begin{eqnarray} \label{e131}
 R_{a} \equiv  \frac{S_{12,3}}{S_{min}} = \frac{b_{12,3}}{b} \epsilon_a 
 \;\;\; , \;\;\; 
 R_{b} \equiv  \frac{S_{12}}{S_{min}} = \frac{b_{12}}{b} \epsilon_b \; ,
\end{eqnarray}
where the dimensionless charge-mass and energy ratios are defined by
Eqs.(\ref{e110}), (\ref{e120}), (\ref{e124}) and
\begin{eqnarray} \label{e140}
 \epsilon_a  \equiv \sqrt{\frac{E}{E-E_{12}}}   \; ,\; 
 \epsilon_b  \equiv \sqrt{\frac{E}{E_{12}}}  \; .
\end{eqnarray}
We first notice that $T>T_{12,3}$ when $R_{a}>1$.  This condition is
determined by only one combinations of masses, charges and energies.
For an energy $E_{12}$ approaching $E$ the ratio $R_{a}$ becomes
infinitely large expressing that then direct decay is preferred
independent of charges and masses. Otherwise we notice that emission
in the first step of a particle with small charge, small mass and high
energy favors the sequential decay mode.  The system would then choose
the best of the stepwise disintegrations through the different
possible intermediate configurations, in all cases provided the
process can compete with the direct decay.

\subsection{Centrifugal barrier}

We now consider the case without Coulomb interaction.  Still we assume
that the effect of the short-range interaction is vanishingly small in
the classically forbidden barrier region.  Then we are left with the
potentials corresponding to the centrifugal barrier, i.e.
\begin{eqnarray} \label{e150}
 V(\rho) &=&  \frac{\hbar^2 (K+3/2)(K+5/2)}{2 m \rho^2} \approx  
 \frac{\hbar^2 (K+2)^2}{2 m \rho^2}  \;  ,
 \\ \label{e160}
 V(r_{13}) &=& E_{12} + \frac{\hbar^2 {\ell}_{12,3} ({\ell}_{12,3}+1)}
 {2 \mu_{12,3} r_{13}^2}  \approx  E_{12} + \frac{\hbar^2 
 ({\ell}_{12,3}+1/2)^2} {2 \mu_{12,3} r_{13}^2}  \;  ,
  \\ \label{e165}
V(r_{12}) &=& \frac{\hbar^2 {\ell}_{12} ({\ell}_{12}+1)}
 {2 \mu_{12} r_{12}^2} \approx \frac{\hbar^2  ({\ell}_{12}+1/2)^2}
 {2 \mu_{12} r_{12}^2} \; , 
\end{eqnarray}
respectively for direct decay and the two steps of the sequential
decay where particle $3$ is emitted first.  We used again both the
appropriate reduced masses and the lowest allowed angular momentum
quantum numbers ${\ell}_{12,3}$ and ${\ell}_{12}$ corresponding to the
relative two-body motion.  The direct decay has the lowest
hypermomentum quantum number $K = {\ell}_{12,3} + {\ell}_{12}$
\cite{nie01}.  More than one partial wave may contribute but in
general the lowest values dominate as they correspond to the lowest
barriers. We inserted the usual improved semiclassical approximation
in Eqs.(\ref{e160}) and (\ref{e165}) for the expectation value of the
angular momentum.

The WKB exponents are then all obtained from the same type of generic
integrals, i.e.
\begin{eqnarray} \label{e170}
 S \propto   \int_{x_0}^{x_t} {\rm d} x \sqrt{\frac{x_t^2}{x^2} - 1}
 &=& -  \sqrt{x_t^2-x_0^2} + x_t \log\Big(\frac{x_t + 
 \sqrt{x_t^2-x_0^2}}{x_0}\Big)   \\ \nonumber
 &\approx&   x_t \log\big(\frac{2 x_t}{e x_0}\big)  \;  ,
\end{eqnarray}
where we assumed $x_0 \ll x_t $, and $ x_t^2 = {\ell} ({\ell} + 1)
\approx ({\ell} + 1/2)^2$.  Furthermore $(x_t/x_0)^2 = E_B/E $ where $E$
is the decay energy and $E_B = \hbar^2 ({\ell} + 1/2)^2 /(2 \mu
R_0^2)$ is the potential energy at the inner radial turning point
$R_0$.  The reduced mass, angular momentum and turning point for the
process in question then has to be inserted.  The WKB transmission
coefficient from Eq.(\ref{e50}) becomes
\begin{equation} \label{e180}
 T = \frac{1}{1 + \Big(\frac{4 E_B}{e^2E}\Big)^{{\ell}+1/2}} \approx 
\Big(\frac{e^2E}{4 E_B}\Big)^{{\ell}+1/2}  \;  ,
\end{equation}
where the last approximation is valid when $E \ll E_B$.  For $s$-waves
we get $T \approx e k R_0$, where $k$ is the wave number associated
with the energy $E$. This agrees with the exact low-energy result, $4
k /K$, for a square well \cite {sie87} when the wave number $K$ inside
the well and the inner classical turning point are related by $K R_0 =
4/e$ which is around 1.5.  Furthermore, for finite ${\ell}$ we also
get the correct low-energy threshold behavior $T \propto
E^{{\ell}+1/2}$ \cite{sie87}.

Using Eqs.(\ref{e150}), (\ref{e160}), (\ref{e165}) and (\ref{e180}) we
then arrive at the WKB transmission coefficients for direct and
two-step sequential decays, i.e.
\begin{eqnarray} \label{e190}
 T = \Big(\frac{2 m \rho_0^2 E} {\hbar^2({\ell}_{12,3} + 
 {\ell}_{12} + 2)^2}\Big)^{{\ell}_{12,3} + {\ell}_{12} + 2} 
 &\approx& \Big(\frac{2 \sum_{i<k} m_im_k R_0^2 E} 
 {M \hbar^2({\ell}_{12,3} + 
 {\ell}_{12} + 2)^2}\Big)^{{\ell}_{12,3} + {\ell}_{12} + 2} \;  ,
 \\ \label{e200}
T_{12,3} =  \Big(\frac{2 \mu_{12,3} R_{12,3}^2 (E-E_{12}) }
{\hbar^2 ({\ell}_{12,3}+1/2)^2}\Big)^{{\ell}_{12,3}+1/2} &\approx&
  \Big(\frac{4 \mu_{12,3} R_{0}^2 (E-E_{12}) }
{3 \hbar^2 ({\ell}_{12,3}+1/2)^2}\Big)^{{\ell}_{12,3}+1/2} \;  ,
 \\ \label{e210}
T_{12} =  \Big(\frac{2 \mu_{12} R_{12}^2 E_{12} }
 {\hbar^2 ({\ell}_{12}+1/2)^2}\Big)^{{\ell}_{12}+1/2} &\approx&
\Big(\frac{2 \mu_{12} R_{0}^2 E_{12} }
 {\hbar^2 ({\ell}_{12}+1/2)^2}\Big)^{{\ell}_{12}+1/2} \;  ,
\end{eqnarray}
where we again used the appropriate reduced masses, angular momenta
and radii for the different processes.  We assumed the initial state
roughly is a triangle with equal side length $R_0$. Furthermore we
expressed $\rho_0$ in Eq.(\ref{e190}) in terms of $R_0$ from an
equation analogous to Eq.(\ref{e20}), see \cite{jen02}.

Thus, again when $T_{12,3} > T$ the sequential decay is dominating and
vice versa.  The exponent $({\ell}_{12,3} + {\ell}_{12} + 2)$ in the
$T$-expression is larger than the sum of the other two exponents
reflecting that a centrifugal barrier exists for the three-body system
even when all relative angular momenta are zero, see Eq.(\ref{e150})
with $K=0$.  Each decay rate increases with increasing reduced mass
and available energy, but first of all with decreasing angular
momentum.
 
Combining Eqs.(\ref{e190}), (\ref{e200}) and (\ref{e140}) we obtain
\begin{eqnarray} \label{e220}
  \frac{T}{T_{12,3}}  \approx
 \Big(\frac{3}{2}\Big)^{{\ell}_{12,3} + 1/2} 
 \Big(\frac{2 E \mu_{12,3} R_0^2}{\hbar^2} \Big)^{{\ell}_{12} + 3/2} 
 \epsilon_a^{2{\ell}_{12,3} + 1} \\ \nonumber
 \Big(\frac{ \sum_{i<k} m_im_k}{M \mu_{12,3}}
 \Big)^{{\ell}_{12,3}+{\ell}_{12} + 2}  
 \frac{ ({\ell}_{12,3} +1/2)^{2{\ell}_{12,3} + 1}}
 {({\ell}_{12,3} + {\ell}_{12} + 2)^{2{\ell}_{12,3} + 2{\ell}_{12} + 4}} \; .
\end{eqnarray}

The $\epsilon_a$ term is always larger than one favoring direct decay.
We see again that for an energy $E_{12}$ approaching $E$, the direct
decay is preferred independent of the masses.  The inverse energy
factor multiplying $E$ is for nuclei of the order of a few MeV$^{-1}$,
which means that $E$ smaller than about $2$~MeV tends to prefer
sequential decay.  The mass average $\sum_{i<k} m_im_k/(M\mu_{12,3})$
is always larger than one and reduces for equal masses to $3/2$.  For
vanishing $m_3$ direct decay is preferred whereas large $m_3$ favors
sequential decay.  Emission of the large mass first is preferred. The
angular momentum factor tends to be much smaller than unity (1/32
already for $s$-waves) and therefore favoring sequential decay.  Then
a finite angular momentum requires a large energy to overcome the
centrifugal barrier.

\subsection{Important examples}

The general expressions and discussions can be much better appreciated
with examples from existing systems. We therefore select a few
particularly important cases.  We first look at three identical
particles of charge $Z_0e$ and mass $A_0m_n$ where $m_n$ for example
is the neutron mass.  For $\alpha$-particles combined in the first
excited $0^+$ state of $^{12}$C where  $E\approx 3 E_{12}$, we get
$b_{12}/b = 0.24$, $b_{12,3}/b = (2/3)^{3/2}$, $\epsilon_a
\approx \sqrt{3/2}$, $\epsilon_b \approx \sqrt{3}$, and therefore $R_{a}
 \approx 2/3$.  Thus sequential decay is preferred in agreement with
\cite{fre94}.

The higher-lying $0^+$-excitation has $E\approx 40 E_{12}$,
$\epsilon_a \approx 1$, $\epsilon_b \approx 6.5$, and therefore $R_{a}
\approx 0.5$. Sequential decay is even more favored. A higher-lying 
intermediate two-body state may also produce sequential decay.  The
$2^+$-state in $^8$Be corresponds to $E \approx 2 E_{12}$, $\epsilon_a
\approx \epsilon_b \approx \sqrt{2}$, and therefore now $R_{a}
\approx 0.7$, still favoring sequential decay through this branch over 
the direct decay.

The ratio of these two sequential transmission coefficients from
Eq.(\ref{e129}) then tremendously favors decay through the two-body
$0^+$-state.  This expectation is in agreement with the analysis of
available experimental data showing that at most 4\% of this decay
occurs without going through the $0^+$ ground state of $^8$Be
\cite{sch66}. However, one complication is that the two $0^+$ states
are coherently populated in experiments.  In principle then both
states produce unseparable contributions in the relevant energy
region.  The attractive two-body interaction may also play a role.

Another example is two protons and a core corresponding to two-proton
decay from proton dripline nuclei.  Two different sequential decays
can occur, i.e. first emission of either the proton or the core. The
relevant mass and charge ratios in the two cases are found to be
$(b_{12}/b, b_{12,3}/b) \approx (0.5/\sqrt{2},0.5/\sqrt{2}),
(0.25/Z_c,1)$, respectively.  Thus sequential proton emission is
preferred over sequential core emission. Direct decay is only favored
when $E_{12}$ is close to $E$, i.e. with the present estimates when
$7 E/8 <E_{12}< E $.

If one proton is replaced by a neutron the specifically favored
two-body structures become decisive. In general we consider one
neutral and two charged particles.  For nuclei the neutron is then one
of the three particles. Let the other two particles have charges
$Z_ce$, $Z_pe$ and masses $A_cm_n$, $A_pm_n$.  There are three possible
sequential decays, i.e. first emission of one of the three
particles. In all three cases one decay is Coulomb dominated and the
other essentially controlled by the centrifugal barriers. The Coulomb
WKB exponents in Eqs.(\ref{e110}) and (\ref{e120}) in general are much
larger than the logarithmic terms in Eq. (\ref{e170}). Therefore the
path with the largest probability is found for the division of
smallest Coulomb $S$-values.  From Eqs.(\ref{e124}), (\ref{e110}) and
(\ref{e120}) we obtain
\begin{eqnarray} \label{e250}
 S_{pc} &=& \frac{\pi}{2} \sqrt{\frac{2 m_n}{\hbar^2}} \frac{Z_p Z_c e^2}
 {\sqrt{E_{pc}}} \sqrt{\frac{A_pA_c}{A_p+A_c}}   \; , \\  \label{e260}
 S_{nc,p} &=& \frac{\pi}{2}  \sqrt{\frac{2 m_n}{\hbar^2}} 
 \frac{Z_p Z_c e^2}{\sqrt{E-E_{nc}}} \sqrt{\frac{(A_c+1)A_p}{A_p+A_c+1}} 
 \; , \\ \label{e270}
 S_{np,c} &=& \frac{\pi}{2} \sqrt{\frac{2 m_n}{\hbar^2}} 
 \frac{Z_p Z_c e^2}{\sqrt{E-E_{np}}} \sqrt{\frac{(A_p+1)A_c}{A_p+A_c+1}} 
\; , \\ \label{e280}
 S &=& \frac{\pi}{2} \sqrt{\frac{2 m_n}{\hbar^2}} 
 \frac{Z_p Z_c e^2}{\sqrt{E}} \sqrt{\frac{A_cA_p}{A_p+A_c+1}} \; . 
\end{eqnarray}
We immediately conclude that all these quantities are rather similar
except for the energy factors.  Thus, both $S_{np,c}$ and $S_{nc,p}$
exceed $S$, i.e. direct decay is always preferred. The heavier core
mass favors sequential emission of the proton over the core. In both
cases high energy emission is favored.

If two neutrons (mass $m_n$) surround a core the centrifugal barriers
are decisive.  Two sequential decays are possible, i.e. neutron or
core emission first.  For a relatively large core mass we get for
neutron emission from Eq. (\ref{e220})
\begin{eqnarray} \label{e290}
  \frac{T}{T_{12,3}}  \approx
 \Big(\frac{4 E m_n R_0^2}{\hbar^2} \Big)^{{\ell}_{12} + 3/2} 
 \Big(\frac{ 3 E }{E-E_{12}}\Big)^{{\ell}_{12,3} + 1/2} \\ \nonumber
 \times \frac{ ({\ell}_{12,3} +1/2)^{2{\ell}_{12,3} + 1}}
 {({\ell}_{12,3} + {\ell}_{12} + 2)^{2{\ell}_{12,3} + 2{\ell}_{12} + 4}} \; .
\end{eqnarray}
For core emission this expression should be divided by
$2^{{\ell}_{12,3} + 1/2}$.  For $s$-waves Eq. (\ref{e290}) is
approximately
\begin{eqnarray} \label{e293}
  \frac{T}{T_{12,3}}  \approx
   \Big(\frac{E}{7 {\rm MeV}}\Big)^{3/2}
 \sqrt{\frac{E}{E - E_{12}}} \; .
\end{eqnarray}
which as always for energies $E_{12}$ very close to $E$ favor direct
decay, but sequential decay is favored for $E_{12}$ relatively close
to $E$ for moderate energies up to a few MeV.

\section{Summary and conclusions}

We investigate the decay of low-lying continuum states into three
particle final states. The initial state may be populated in reactions
or by other decays like beta-decay.  We assume that the decay
mechanism is independent of how the initial state was formed.  We
consider the partial decay of a many-body resonance state into three
specific fragments.  We assume that these three fragments are formed
with a certain probability before they enter regions of larger spatial
extension on their way to total separation as observed in the final
state.  The analogy is $\alpha$-decay where a preformation factor
describes the probability that the $\alpha$-particle exists before its
attempts to penetrate the barrier. This relates to definitions of
three-body spectroscopic factors.

We choose the hyperradius as the adiabatic coordinate which is an
average radial coordinate obtained as a mass weighted mean square
average of distances between pairs of particles. We can then study the
structure along the path defined by any of the adiabatic potentials.
Then one particle may increase its distance to the two other particles
which first choose to stay together and at some later point also
separate. This is naturally called sequential decay. The system may
also prefer to increase all pairwise distances simultaneously. This is
called direct decay.  Thus we can study the conditions for choosing
the different decay mechanisms.

We use a schematic model with only Coulomb and centrifugal barrier
terms effective at distances larger than a minimum hyperradius.  We
discuss how to classify and characterize the sequential and direct
decay mechanisms.  With different geometric configurations we compare
analytically the probabilities for sequential and direct decays. When
all three particles are charged the Coulomb potential dominates over
the centrifugal barrier. Sequential decay by emission of a high energy
particle of small mass and small charge is most favorable simply
because the corresponding barrier then is smaller.  The short-range
interaction are expected to provide the intermediate stepping stone of
a favorable configuration like a two-body resonance or perhaps only by
exploiting an attraction efficient at short distances.  When only a
large centrifugal barrier term is present sequential decay by emission
of the large mass seems to be most likely, but if this barrier is
relatively small the intermediate configuration cannot be reached
before the full decay has taken place.  These conclusions can be
modified or changed by inclusion of the short-range interaction as
discussed in the following companion paper.


\begin{thebibliography}{99}

\bibitem{sie87} P.J. Siemens and A.S. Jensen, 
Elements of nuclei. Many-body physics with the strong interaction, 
Addison-Wesley, California 1987.

\bibitem{goo65} Proc. Conf on Correlations of Particles Emitted in 
Nuclear Reactions, ed. C.D Goodman, Rev. Mod. Phys. {\bf 37} (1965) 327.

\bibitem{oza01} A. Ozawa, T. Suzuki and I. Tanihata,  
Nucl. Phys. {\bf A 693} (2001) 32.

\bibitem{jen04} A.S.~Jensen, K. Riisager, D.V.~Fedorov and E. Garrido,
Rev. Mod. Phys. {\bf 76} (2004) 215.

\bibitem{kar04} O.I. Kartavtsev, Few-Body Systems {\bf 34} (2004) 39.

\bibitem{dan87} B.V. Danilin, M.V. Zhukov, A.A. Korsheninnikov, L.V. Chulkov 
V.D. Efros, Sov. J. Nucl. Phys. {\bf 46} (1987) 225.

\bibitem{boc93}  O.V. Bochkarev et al., 
Izv. Akad. Nauk. Fiz. {\bf  57} (1993) 183.



\bibitem{gar05} E. Garrido, D.V. Fedorov, A.S.~Jensen and H.O.U. Fynbo,
submitted to Nucl. Phys. {\bf A }.

\bibitem{ich73} M. Ichimura and A. Arima, Nucl. Phys. {\bf A 204} (1973) 225.

\bibitem{nie01} E. Nielsen, D.V. Fedorov, A.S. Jensen and E. Garrido, 
Phys. Rep. {\bf 347} (2001) 373.

\bibitem{fed94} D.V. Fedorov, A.S. Jensen and K. Riisager,
Phys. Rev. {\bf C 50} (1994) 2372.

\bibitem{fed02}  D.~V.~Fedorov, A.~S.~Jensen and H.O.U. Fynbo,
Nucl. Phys. {\bf A 718} (2003) 685c.

\bibitem{gar96}  E. Garrido, D.V. Fedorov and A.S. Jensen,
Phys. Rev. {\bf C 53} (1996) 3159.

\bibitem{jen02} A.S. Jensen, K. Riisager, D.V. Fedorov and E. Garrido, 
Europhysics Lett. {\bf 61} (2003) 320.

\bibitem{fre94} M. Freer et al., Phys. Rev. {\bf C 49} (1994) 1751.

\bibitem{sch66} D.~Schwalm and B.~Povh, Nucl. Phys. {\bf 89} (1966) 401.


\end{thebibliography}
\end{document}